 \documentclass[aps,prc,twocolumn,showpacs,floatfix,4paper]{revtex4-1}
\usepackage{amsmath}
\usepackage{amssymb}
\usepackage{epsfig}
\usepackage{graphicx}
\def\bfg #1{{\mbox{\boldmath $#1$}}}

\begin{document}
\title{Polarized proton-deuteron scattering as a test of time-reversal invariance}

\author {Yu.N. Uzikov$^{1,2,3}$, J. Haidenbauer$^4$ }

\affiliation{
$^1$Laboratory of Nuclear Problems, Joint Institute for Nuclear
Research, Dubna, 141980 Russia\\
$^2$Department of Physics, Moscow State University,  Moscow, 119991  Russia\\
$^3$ Dubna State University, Dubna, 141980 Russia\\
$^4$Institute for Advanced Simulation and Institut f\"ur Kernphysik, 
Forschungszentrum J\"ulich GmbH, D-52428 J\"ulich, Germany
}

\begin{abstract}
Scattering of protons with transversal polarization $p_y^p$ on deuterons 
with tensor polarization $P_{xz}$ provides a null-test signal
for time-reversal (T) invariance violating but parity (P) conserving effects. 
We calculate the corresponding null-test observable at beam energies $100$--$1000$ MeV 
within the spin-dependent Glauber theory considering T-violating P-conserving nucleon-nucleon interactions.
The $S$-wave component of the deuteron wave function as well as the $D$-wave
are taken into account and the latter is found to play an important role for the 
magnitude and the energy dependence of the observable in question.
Specifically, with inclusion of the $D$ wave the maximum of the obtained signal is 
shifted to higher beam energies, i.e. to $700-800$ MeV. 
\end{abstract}

\pacs{24.80.+y, 25.10.+s, 11.30.Er, 13.75.Cs}
\maketitle
\section{Introduction}
\label{intro}

The discrete symmetries of parity (P) and time reversal (T) play a crucial role in our understanding
of fundamental interactions. For example, P-violation led to the discovery of the
V-A structure of the weak interaction of leptons and quarks.
CP-violation (or T-violation assuming CPT symmetry, where C stands for charge conjugation) 
is required to account for the baryon asymmetry of the universe \cite{sakharov}. 
Since within the standard model the CP-violation observed in physics of kaons
and $B$-mesons is by far not sufficient to explain 
this asymmetry, other sources of CP-violation have to be found.
The possibility that T-violation might arise due to physics beyond the standard model 
 has motivated a number of low-energy experiments. These measurements are classified according to
 whether or not the measured observables violate parity as well as time reversal invariance. 
 Various efforts are undertaken with respect to measurements of the electric dipole moments (EDM) 
 of elementary particles and atoms which are both T-violating and P-violating
 observables. The other type of observables which are T-violating but 
 P-conserving (TVPC) received considerably less attention. These include tests of
 the detailed balance \cite{blanke,french},
 correlations in beta-decay of polarized neutrons \cite{y}, charge-symmetry breaking
 in $pn$ scattering \cite{simonius97}, and
  transmission of polarized neutrons through tensor-polarized nuclei \cite{huffman}.
  The reason why TVPC observables are interesting is that the experimental limits on 
  them are still quite weak, in particular much weaker as compared to the EDM.
Theoretically, effects of TVPC physics beyond the standard model 
can be studied in a model-independent way using effective field theory. 
In this context it was shown that the limits on the EDM imply also severe bounds on TVPC 
observables \cite{conti-khripl}.
   Nonetheless, exceptions to this may occur depending on either unknown details of
   the generation of the EDM within specific scenarios beyond the standard model \cite{kurilov}
   or on the possible existence of right-handed neutrinos \cite{menuofif2016}.
  
In the present paper we consider the scattering of protons with transversal polarization 
$p_y^p$ on deuterons with tensor polarization $P_{xz}$.
This double-polarized proton-deuteron ($pd$) scattering process allows access to a 
null-test observable for TVPC effects~\cite{conzett}.
By definition such observables are non-zero only in the presence of a TVPC interaction 
and cannot be generated by the T-invariant initial or final state interaction.
 The observable in question is the total (integrated) cross section for that scattering 
reaction and it will be called ${\widetilde \sigma}$ in the following. 
 An experiment to measure this quantity is planned at the COSY accelerator in J\"ulich \cite{TRIC},
at a projected laboratory energy of $135$ MeV. The first theoretical analysis of this observable
 was performed in Ref.~\cite{beyer} in a calculation of the nonmesonic
 breakup of the deuteron within the single scattering approximation for some type of TVPC forces.
 The above energy was found to be the
 most sensitive one to the TVPC effects. Later on Faddeev calculation were performed for 
 neutron-deuteron ($nd$) scattering, but at much lower energies namely $100$~keV \cite{Lazauskas}.
 Recently, in Ref.~\cite{TUZyaf,UzTePRC} the generalized optical theorem was applied 
 to calculate the cross section ${\widetilde \sigma}$  
 at proton beam energies of $T_p=100$--$1000$ MeV. Here the spin-dependent Glauber theory \cite{PK}
 was used to get the forward elastic $pd$ scattering amplitude.  
 It was shown \cite{UzTePRC} that the double-scattering mechanism, ignored in \cite{beyer}, 
 dominates the null-test observable ${\widetilde \sigma}$. The Coulomb interaction was taken 
 into account and found to lead to no divergence of this observable.
 As in Ref.~\cite{beyer}, lower energies of around $100$ MeV were found to be more preferable 
to search for a TVPC signal \cite{UzTePRC}.
 ``Null combinations'' of some differential spin observables of $pd$ elastic scattering, i.e. 
quantities which deviate from zero only in case of the presence 
 of TVPC effects, were analysed in Refs.~\cite{TUZizv,TUZizv16}.
  
One approximation made in Ref.~\cite{UzTePRC} is that 
only the $S$-wave component of the deuteron wave function was taken into account.
Terms which include the $D$-wave, somewhat cumbersome to calculate, were neglected.
Therefore, in the present work, now special attention is paid to the role played by the 
deuteron $D$-wave for this null-test observable.
Indeed, the prime reason why the contribution of the $D$-wave was not studied in \cite{UzTePRC}
is the analysis (of elastic $pd$ scattering) in Ref.~\cite{PK}, where it was argued 
that at energies $\sim 100$ MeV
its contribution is less important than the $S$-wave contribution, assuming in particular that 
at zero transferred 3-momentum the contribution of $D$-wave vanishes.  
Here we account for the deuteron $D$-wave and we employ the same TVPC interactions 
again as in \cite{beyer,UzTePRC}. We show that the $D$-wave is very important for the
absolute value of the null-test signal ${\widetilde \sigma}$ and its energy dependence.
Preliminary results of the present study were presented at conferences \cite{Uzdspin15,UzFB15}.

The paper is organized as follows: In Sec. II we outline the employed formalism. 
Our numerical results are reported in Sec. III. Finally, in Sect. IV a summary and
some conclusions are presented. 

\section{Null-test signal of TVPC forces}

The total cross section of $pd$ scattering for the case of a 
T-violating P-conserving $NN$ interaction can be written as~\cite{SPIN2014,UzTePRC}
\begin{eqnarray}
\label{totalspin}
\sigma_{tot}=\sigma_0&+&\sigma_1 {\bf p}^{ p}\cdot {\bf p}^d +
\sigma_2 ({\bf p}^{ p}\cdot { {\bf m}}) ({\bf p}^d\cdot { {\bf m}}) \nonumber  \\
&+& \sigma_3 { P_{zz}} + {\widetilde \sigma}\, {p_y^p P_{xz}^d},
\end{eqnarray}
where ${\bf p}^p$ (${\bf p}^d$) is the vector polarization of the initial proton (deuteron) and
$P_{zz}$ and $P_{xz}$ are the tensor polarizations of the deuteron. The OZ axis is directed along
the proton beam momentum ${\bf m}$, OY$\uparrow\uparrow {\bf p}^p$,
OX $\uparrow\uparrow [{\bf p}^p\times {{\bf m}}]$.
%
%
 In Eq.~(\ref{totalspin}) the terms $\sigma_i$ with $i=0,1,2,3$ are non-zero only
 for T-invariant P-conserving interactions and the last term ${\widetilde \sigma}$
constitutes a null-test signal of T-violation with P-conservation.
 In the notation of Ref.~\cite{UzTePRC} the integrated cross section ${\widetilde \sigma}$
 is related to the TVPC term $\widetilde g$ in the forward $pd$ elastic scattering amplitude by 
\begin{eqnarray}
\label{sigma5}
{\widetilde\sigma}=-4\sqrt{\pi}\frac{2}{3}{\rm Im}\,{\widetilde g}.
\end{eqnarray}
 
In order to calculate the TVPC amplitude $\widetilde g$ we use the Glauber theory of multistep scattering.
In addition to the regular (T-invariant) $NN$ scattering amplitudes \cite{arndt} we consider 
the following terms of the TVPC $NN$ interaction which were under discussion in Refs.~\cite{beyer,UzTePRC}:
\begin{widetext}
\begin{eqnarray}
\label{TVNN}
t_{pN}&=&{h_N[({\bfg \sigma} \cdot {\bf k})({\bfg \sigma}_N \cdot {\bf q})+
({\bfg \sigma}_N \cdot {\bf k})({\bfg \sigma} \cdot {\bf q})-
\frac{2}{3}({\bfg \sigma}_N \cdot{\bfg \sigma})
({\bf k}\cdot {\bf q}) ]}/m_p^2
 \nonumber \\
&+&g_N [{\bfg \sigma} \times {\bfg \sigma}_N]\cdot [{\bf q }\times{\bf k}][{\bfg \tau} -{\bfg \tau}_N]_z/m_p^2
+{g^\prime_N ({\bfg \sigma} - {\bfg \sigma}_N)\cdot i\,[{\bf q}\times {\bf k}]
[{\bfg \tau} \times{\bfg \tau}_N]_z}/m_p^2.
\end{eqnarray}
\end{widetext}
Here ${\bfg \sigma}$
(${\bfg \sigma}_N$)
 is the Pauli matrix acting on the spin state of the proton
(nucleon $N=p,n$),
${\bfg \tau}$
(${\bfg \tau}_N$)
is the corresponding matrix acting on the isospin state, and $m_p$ is the proton mass.
The transferred and average momenta, $\bf q$ and $\bf k$, are defined in terms of
the final ($\bf  p'$) and initial ($\bf  p$) center-of-mass momenta of the nucleons by
${\bf q}=({\bf p}- {\bf  p'})$ and ${\bf k}=({\bf p}+ {\bf  p'})$. 
In the framework of phenomenological meson-exchange interactions the term $g^\prime$ 
results from $\rho$-meson exchange, while 
the $h$-term comes from the exchange of the axial-vector meson $h_1$ with quantum numbers 
$I^G(J^{PC})=0^-(1^{+-})$.
Up to now, no definite interpretation of $g_N$ in terms of meson exchanges has been 
given, but its contribution is considered here following Refs.~\cite{beyer,UzTePRC}. Note, however, 
that in the present study we take into account that there is no contribution of the $g$-term 
to $pp$ scattering because of the Pauli principle \cite{herczeg,bystricky}

As was shown in \cite{UzTePRC}, with the TVPC amplitudes in Eq.~(\ref{TVNN}) only the
double-scattering mechanism gives a contribution to
the null-test signal, whereas the single-scattering mechanism does not. 
The double-scattering amplitude is given by the following integral 
\begin{eqnarray}
\label{DS}
M^{(d)}=\frac{i}{2\pi^{3/2}}\iint d^2q'M({\bf q},{\bf q}'; {\bf S}, {\bfg \sigma}) \ ,
\end{eqnarray}
where $M({\bf q},{\bf q}'; {\bf S}, {\bfg \sigma})$ is the Glauber operator of $pd$ scattering.
In order to get the TVPC amplitude $\widetilde g$ one has to calculate the expectation
value of this operator 
for definite initial $|\mu,\lambda\rangle$ and final $|\mu',\lambda'\rangle$ spin states 
at $\bf q=0$:
\begin{widetext}
\begin{eqnarray}
{\widetilde g}=\frac{1}{(2\pi)^{3/2}}\int d^2 q^\prime
 \langle\mu'=\frac{1}{2},\lambda'=0|M({\bf q}=0,{\bf q}^\prime; {\bf S}, {\bfg \sigma})|
\mu=-\frac{1}{2},\lambda=1\rangle,
\label{g5ds}
\end{eqnarray}
\end{widetext}
where $\mu (\mu')$ and $\lambda (\lambda')$  are the spin projections of the initial (final) proton
and the deuteron on the quantization axis.

As found in \cite{UzTePRC}, the $g^\prime$ term in Eq.~(\ref{TVNN}) gives zero contribution 
to the amplitude ${\widetilde g}$, for both the $S$ and $D$ waves of the deuteron. 
Some qualitative arguments for this result are discussed in Ref.~\cite{UzFB15}.
For the $h$- and $g$-terms of the TVPC $pN$ interaction one has from 
Eq.~(29) of Ref.~\cite{UzTePRC}:
\begin{widetext}
\begin{eqnarray}
\label{18total}
M({\bf q}, {\bf Q}; {\bf S}, {\bfg \sigma})=
 {W_{ij}}\left\{ {S}_i,{S_j} \right\} S_{0}^{\left( 0 \right)}
  -\sqrt{2}\, W_{ij}\,\left[ \left \{ {S}_i,S_j \right \}S_{12}(
  \hat{\bf Q};{\bf S},{\bf S}  )
  +{{S}_{12}}( \hat{\bf Q};{\bf S},{\bf S} )\left\{ {{S}_{i}},{{S}_{j}} \right\} \right ]
S_{2}^{\left( 1 \right)}+ \nonumber \\
+\frac{1}{16\pi }{{W}_{ij}}\int{{{d}^{3}}r
\frac{1}{r^2}{{e}^{i{\bf Q}{\bf r}}}{{w }^{2}}}\,
  {{S}_{12}}\left( \hat{\bf r};{{{\bfg \sigma }}}_{n},
{{{\bfg{\sigma }}}_{p}} \right)\left\{ {{S}_{i}},{{S}_{j}}
 \right\}{{S}_{12}}\left( {\bf \hat{r}};{{{\bfg \sigma }}_{n}},{{{\bfg \sigma }}_{p}} \right).
\end{eqnarray}
\end{widetext}

 Here
\begin{eqnarray}
\label{T2}
 S_{12}({\hat{\bf r}};{\bfg\sigma}_p,{\bfg\sigma}_n)
=3({\bfg\sigma}_p\cdot {\hat{\bf r}})({\bfg\sigma}_n\cdot {\hat{\bf r}})-
{\bfg\sigma}_p\cdot{\bfg\sigma}_n
 \end{eqnarray}
 is the tensor operator,
  ${\bfg\sigma}_n ({\bfg\sigma}_p)$ are the Pauli matrices acting on the spin states of the neutron
 and proton in the deuteron, and ${\hat{\bf r}}$  is the unit vector
  directed along the radius-vector ${\bf r}$.
  We use the notations $\{S_i,S_j\}=S_iS_j+S_jS_i$,
 where ${\bf S}=({\bfg \sigma}_n+{\bfg \sigma}_p)/2$. 
The tensor operator $S_{12}({\hat {\bf Q}}; {\bf S}, {\bf S})$ 
is defined analogous to Eq.~(\ref{T2}).
 
The deuteron form factors appearing in Eq.~(\ref{18total})
are related to the $S$- and $D$-wave components of the
deuteron wave function, $u$ and $w$ \cite{PK}:
\begin{eqnarray}
\label{ssds}
S_0^{(0)}(q)&=&\int_0^\infty dr\,u^2(r)j_0(qr), \nonumber \\
S_0^{(2)}(q)&=&\int_0^\infty dr\,w^2(r)j_0(qr), \nonumber \\
S_2^{(1)}(q)&=&2\int_0^\infty dr\,u(r)w(r)j_2(qr), \\
S_2^{(2)}(q)&=&-\frac{1}{\sqrt{2}}\int_0^\infty dr\,w^2(r)j_2(qr), \nonumber \\
S_1^{(2)}(q)&=&\int _0^\infty dr\,w^2(r) j_1(qr)/(qr) . \nonumber
\end{eqnarray}
In Eq.~(\ref{18total}) the summation has to be done over recurring indices $i,j=x,y,z$.
To perform the integration over the directions of the vector $\bf r$ in Eq.~(\ref{18total}),
we use the following relation \cite{faldt}
\begin{eqnarray}
\iint d\Omega_{\bf r}\exp{(-i\bf Q\bf r)}T_l(\hat {\bf r})= 4\pi j_l(Qr) (-i)^l T_l(\hat {\bf Q}), \nonumber \\
\label{faeld}
\end{eqnarray}
where $j_l(x)$ is the spherical Bessel function,
$T_2(\hat {\bf n})={(\bfg \sigma_p\cdot \hat {\bf n})}(\bfg \sigma_n\cdot \hat {\bf n})-
\frac{1}{3}(\bfg \sigma_p\cdot\bfg \sigma_n)$, $T_0(\hat {\bf n})=\bfg \sigma_p\cdot \bfg \sigma_n$;
$\hat {\bf n}$, $\hat {\bf Q}$, and $\hat {\bf r}$ are unit vectors along ${\bf n}$, ${\bf Q}$ and
$ {\bf r}$, respectively.

The tensor operator $W_{ij}({\bfg \sigma})$ in Eq.~(\ref{18total}) acts only on the spin state of 
the beam proton and does not depend on 
the spins and coordinates ${\bf r}$ of the target nucleons. The explicit expressions for $W_{ij}$ are given in
Ref.~\cite{UzTePRC} for the $h$- and $g$-terms. These operators contain products of one TVPC amplitude 
($g_N$ or $h_N$) with the (T-invariant) $NN$ amplitude $C^\prime_N$ \cite{PK}, namely $C^\prime_nh_p$ 
and $C^\prime_ph_n$ for the $h$-term, and $C^\prime_pg_n$ for the $g$-term. 
The other hadronic amplitudes $A_N$, $C_N$, $B_N$, $G_N$, $H_N$ in the notation of \cite{PK} 
do not contribute to the amplitude ${\widetilde g}$ with regard to the $h$- and $g$-terms.

As already mentioned, the $g'$-term does not contribute to the 
 null-test observable ${\widetilde\sigma}$ within the Glauber theory of $pd$ elastic
 scattering \cite{UzTePRC,UzFB15}. Considering the $h$- and $g$-terms for the double-scattering mechanism 
 and taking into account both the $S$- and $D$-wave of the deuteron we find the following result
 for the forward TVPC amplitude:
\begin{widetext}
 \begin{eqnarray}
\label{g5}
{\widetilde g}=\frac{i}{4{\pi}m_p}
\int_0^\infty dq q^2\left[S_0^{(0)}(q)-\sqrt{8} S_2^{(1)}(q) -4 S_0^{(2)}(q)+ \sqrt{2}\frac{4}{3} S_2^{(2)}(q)+
9 S_1^{(2)}(q)\right]
[-C^\prime_n(q)\,h_p +C^\prime_p(q)(g_n-h_n)], \nonumber \\
 \end{eqnarray}
\end{widetext}
where $S_i^{(j)}$ are the elastic form factors of the deuteron defined in Eq.~(\ref{ssds}).
(Note, however, that one of the form factor, $S_1^{(2)}(q)$, is absent in the electromagnetic 
structure of the deuteron.) In the calculation presented in Ref.~\cite{UzTePRC}
only the first term in the (big) square brackets in Eq.~(\ref{g5}), $S_0^{(0)}(q)$, was taken into account. 
This corresponds to the $S$-wave approximation. The second term,
$S_2^{(1)}(q)$, results from the interference of the $S$- and $D$-state wave functions
while the last three terms contain pure $D$-wave contributions.
One can see from Eq.~(\ref{g5}) that ${\widetilde g}$ contains only products of 
the T-invariant $NN$ amplitude and the TVPC $NN$ amplitude. 
This means that any T-invariant P-conserving background is excluded from the 
null-test observable ${\widetilde\sigma}$. Accordingly, small remaining 
uncertainties in the regular $NN$ scattering amplitudes cannot affect the final 
result for ${\widetilde\sigma}$ significantly. 

\begin{figure}[htb]
\vspace{0.3cm}
\includegraphics[width=0.45\textwidth,clip]{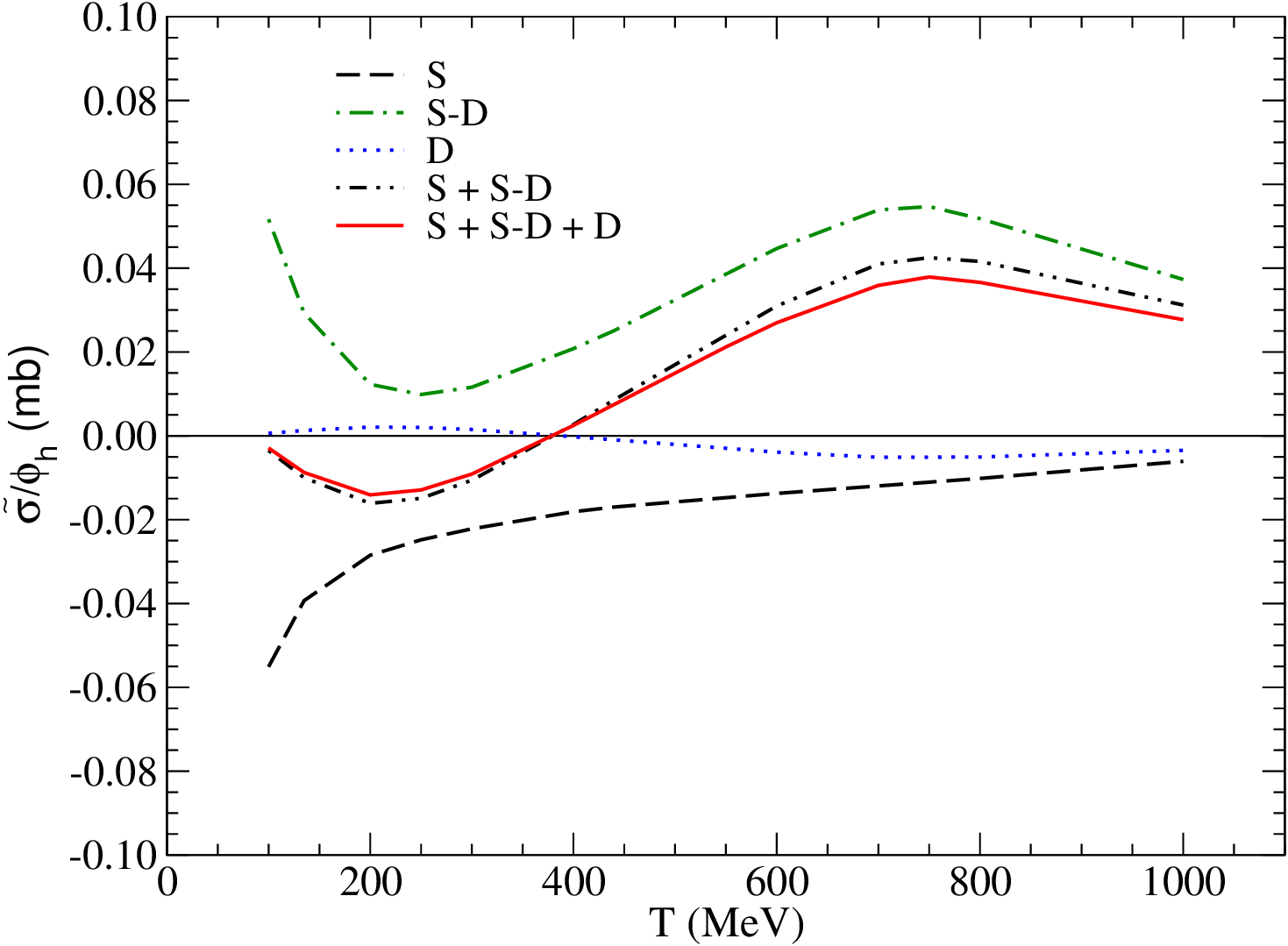}
\caption{Dependence of the TVPC signal $\widetilde\sigma$ 
on the proton beam energy $T$ for the $h$-term, in units of the unknown 
ratio ($\phi_h$) of the TVPC and the strong $h_1NN$ coupling constants.
Results are presented for the different  
contributions due to the $S$- and $D$-wave components of the 
deuteron wave function (CD Bonn), according to the terms in Eq.~(\ref{g5}). 
$S$-wave (dashed line), 
$S$-$D$ interference (dash-dotted line),
$D$-wave (dotted line),
$S$-wave + $S$-$D$ interference (dash-double dotted line), 
total result (solid line). 
}
\label{fig:fig1}
\end{figure}

\section{Results}
In our numerical calculations we employ the $pN$ scattering amplitude $C^\prime_N$
evaluated from the SAID partial wave analysis \cite{arndt}.
The deuteron wave function is taken from the CD Bonn $NN$ model \cite{CDBonn}. 
The amplitude $h_N$ is generated from the exchange of the axial-vector meson $h_1$(1170). 
For explicit expressions of the potential and the resulting amplitude see Eqs.~(23)
--(24) in Ref.~\cite{UzTePRC}. 
The results of the calculation of the observable ${\widetilde\sigma}$ based on the 
$h$-term are presented in Fig.~\ref{fig:fig1}, in units of the unknown ratio 
$\phi_h={\bar G}_h/G_h$, where ${\bar G}_h$ is the TVPC coupling constant and $G_h$ 
the strong coupling constant of the $h_1$ meson with the nucleon.
We employ the very same amplitude as in Ref.~\cite{UzTePRC} so that our present 
result in the $S$-wave approximation coincides with the one given in that reference.
(Note, however, that in Fig.~3 of \cite{UzTePRC} $|{\widetilde\sigma}/\phi_h|$ is shown.)
 
\begin{figure}[htb]
\vspace{0.1cm}
\includegraphics[width=0.45\textwidth,clip]{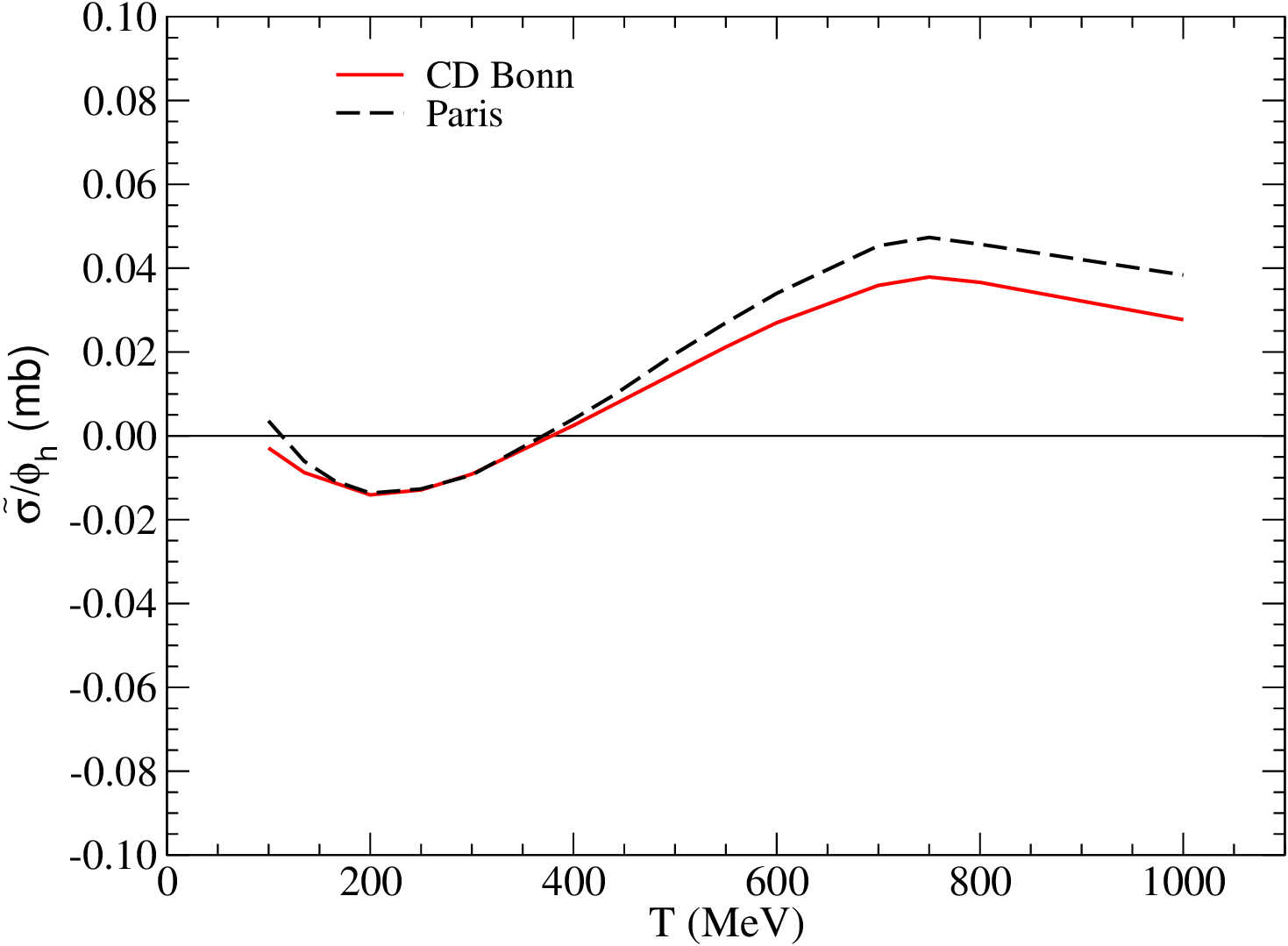}
\caption{Dependence of the TVPC signal $\widetilde\sigma$ 
on the proton beam energy $T$ for the $h$-term, in units of the 
unknown ratio ($\phi_h$) of the TVPC and the strong $h_1NN$ coupling constants.
Total result based on the CD Bonn (solid) and Paris (dashed) deuteron wave
functions are presented.  
}
\label{fig:fig1a}
\end{figure}

From Fig.~\ref{fig:fig1} one can see that the $D$-wave component of the 
deuteron, taken into account in the present
study, has a strong impact. It changes the result for the observable ${\widetilde\sigma}$ 
considerably as compared to the $S$-wave calculation published previously \cite{UzTePRC}. 
It turns out that the $S$-$D$ interference (second term in Eq.~(\ref{g5}) is destructive 
with respect to the pure $S$-wave contribution. As a consequence, it drastically reduces 
the null-test signal ($\widetilde \sigma$) at energies $\sim$100 MeV as compared to
the pure $S$-wave contribution, i.e. in the region of the planned COSY 
experiment \cite{TRIC}. At the same time the interference term provides an enhancement 
of the signal at $700$--$800$ MeV. 
The effect of the pure $D$-wave (last three terms in Eq.~(\ref{g5})), on the other hand,
is indeed negligible, in agreement with what was assumed in Ref.~\cite{UzTePRC}.
Since the energy dependence of ${\widetilde\sigma}$ turned out to be rather sensitive to 
the $D$-wave component we performed calculations with another deuteron wave function, 
namely the one of the Paris $NN$ potential \cite{Paris}. Corresponding results are
presented in Fig.~\ref{fig:fig1a}. Obviously, there are variations
on the quantitative level but qualitatively the resulting energy dependence of 
$\widetilde \sigma$ is similar. 

\begin{figure}[htb]
\vspace{0.1cm}
\includegraphics[width=0.45\textwidth,clip]{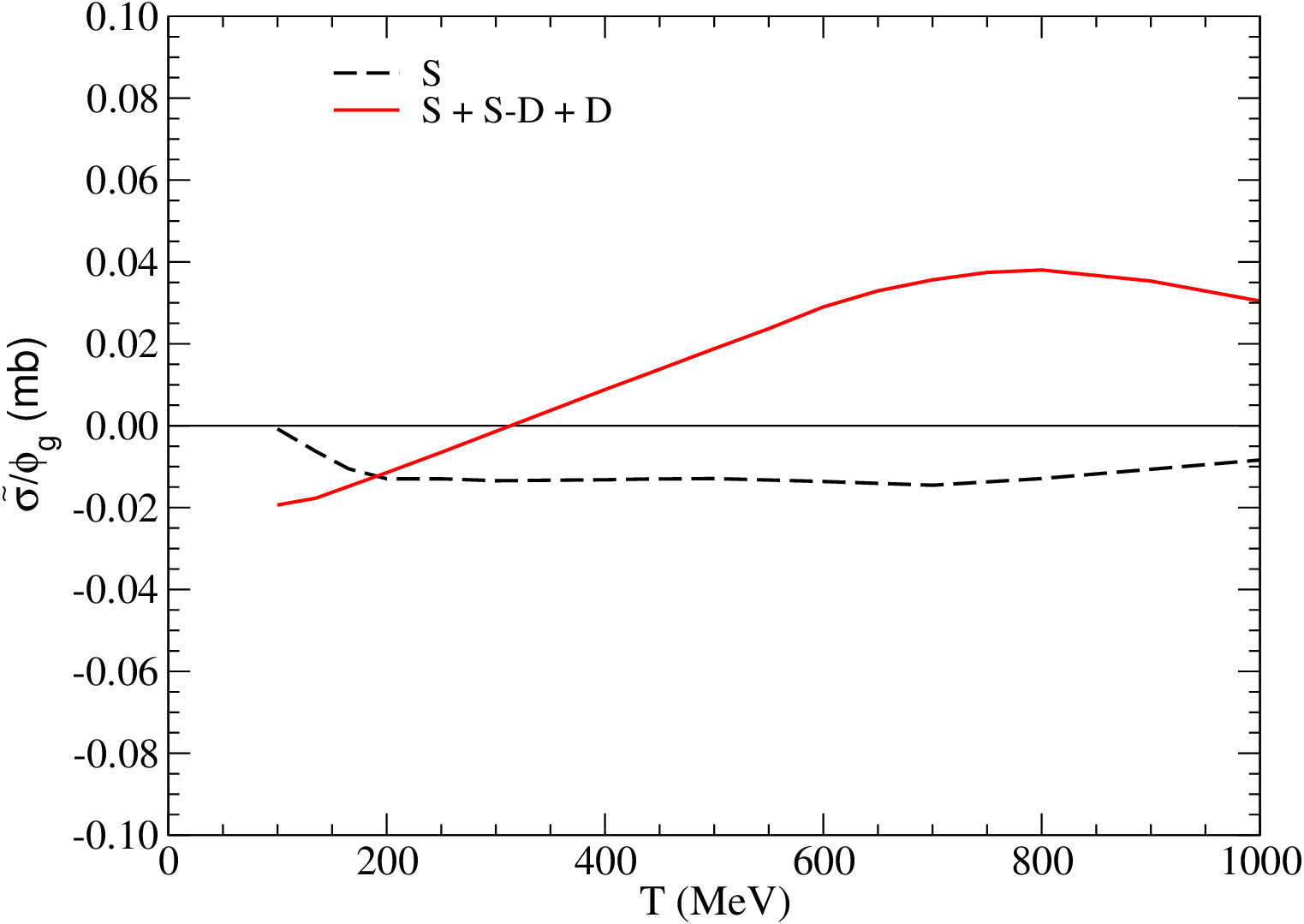}
\caption{Dependence of the TVPC signal $\widetilde\sigma$ 
on the proton beam energy $T$ for the $g$-term. 
Results based on the $S$-wave component of the (CD Bonn) deuteron wave function
(dashed line) and the full results (solid line) are presented.
For the definition of the scaling factor $\phi_g$, see text. 
}
\label{fig:fig2}
\end{figure}

As said above, no definite interpretation of the $g$-term in terms of meson exchanges 
has been given in the literature (see, however, the pertinent remarks in Ref.~\cite{simonius75}). 
Based on the expectation that it originates from very short-ranged dynamics 
the corresponding amplitude was assumed to be constant in Refs.~\cite{beyer,UzTePRC}.
Then the energy dependence of ${\widetilde\sigma}$ is solely determined by that of
the T-invariant $NN$ amplitude $C'_p$, see Eq.~(\ref{g5}). 
Since $h_1$ exchange is also fairly short-ranged it leads to a TVPC amplitude 
that is likewise practically constant \cite{UzTePRC}. On the other hand, the $h_1$ is 
an isoscalar meson and, accordingly, the amplitude $\tilde g$ results from 
the sum of $C'_p$ and $C'_n$. Despite of that the $g$- and $h$-terms yield a fairly 
similar energy dependence of ${\widetilde\sigma}$, as can be seen by comparing the 
results in Figs.~\ref{fig:fig1} and \ref{fig:fig2}. 
(In order to facilite the comparison we scaled the results based on the $g$-term by 
a factor $\phi_g$ so that at the maximum the total results are roughly the same.)
Note that for $h$-type contributions a variety of isospin structures is possible
\cite{herczeg,simonius75} so that, in principle, any combination of $C'_p$ and $C'_n$
in Eq.~(\ref{g5}) can occur. 

\section{Conclusions}
  
The generalized optical theorem was used in Ref.~\cite{UzTePRC} for a calculation of 
$\widetilde \sigma$, the null-test signal for time-reversal violating parity-conserving
effects in proton-deuteron scattering. In this case only the evaluation of the forward 
elastic $pd$ scattering amplitude is required. 
It was found within the Glauber theory that the single-scattering 
mechanism leads automatically to the result {\it zero} for $\widetilde \sigma$. 
Incidentally, it was 
exactly this mechanism that was used in the first theoretical analysis of the null-test 
observable $\widetilde \sigma$ performed in Ref.~\cite{beyer}
via  a straightforward calculation of inelastic and elastic $pd$ scattering.
The double-scattering mechanism, based on the $h$ and $g$ terms of the TVPC $NN$ amplitude,
yields several contributions to the TVPC amplitude of $pd$ scattering, 
but it turned out that it is only one hadronic (T-invariant) $pN$ amplitude, 
namely $C_N^\prime$, that modulates the TVPC observable $\widetilde \sigma$ \cite{UzTePRC}.
   
In the present study we extended our previous investigation \cite{UzTePRC} by taking 
into account the $D$-wave component of the deuteron. We showed for the case of the $h$-
and $g$-type interactions that the deuteron $D$-wave has a strong impact on the 
null-test signal, due to contributions that arise from the interference between 
the deuteron $S$- and $D$-state wave functions. The effect of the $D$-wave component 
alone turned out to be negligible.
Evidently, with the $D$-wave included, a zero crossing of $\tilde\sigma$ is possible 
even when the TVPC interaction itself is non-zero. In the present calculation
this occurs at lower energies, i.e. below $T=400$ MeV. Thus, it is advisable 
to perform experiments at two or possibly more energies in order to achieve conclusive 
results. In any case, our predictions suggest that energies around $700-800$ MeV could be 
more promising for finding a signal. Though there is also some sensitivity to variations 
in the $D$-wave component the overall modification is not too dramatic. 
  
The $g^\prime$-term caused by the $\rho$-meson exchange in the TVPC $NN$ interaction
does not contribute to $\widetilde \sigma$ within the Glauber theory
with T-invariant P-conserving $NN$ interactions in the deuteron \cite{UzTePRC}. 
One-pion exchange is excluded from the TVPC $NN$ interaction from the 
beginning \cite{simonius97}. 
There are several other TVPC terms in the $NN$ interaction 
\cite{herczeg} we ignored here, but which one could examine in the future. 
It should be said, however, that most of those vanish on-shell and, therefore, are presumably
suppressed.  Indeed, a recent study of the P-violating $NN$ interaction within chiral 
effective field theory, where terms that contribute only off-shell arise as well 
\cite{deVries2}, found such contributions to be negligible, at least for 
the energies considered. In any case, off-shell terms do not contribute in 
the Glauber theory of $pd$ elastic scattering.
 
The remaining terms listed in Ref.~\cite{herczeg} are all of $h$-type spin-momentum structure
and differ only in their isospin dependence. For example, on the meson-exchange 
level contributions of isovector nature result from the sometimes considered axial-vector 
meson $a_1$(1260) \cite{beyer,simonius75}. 
But since the spin-momentum structure is the same, any such additional TVPC $NN$ 
interactions can lead only to a modification of Eq.~(\ref{g5}) with regard to the relative 
weight of the $C'_p$ and $C'_n$ amplitudes, but there is no change in the combination 
of the deuteron form factors. 

\vspace{0.3cm}
\begin{acknowledgements}
We thank Jordy de Vries for useful discussions.
This work was supported in part by the Heisenberg-Landau program.
\end{acknowledgements}

\end{document}